# Mikhail Vasil'evich Lomonosov

## *Oration on Benefits of Chemistry* [1]

*read by Mikhail Lomonosov at the public assembly of the Imperial Academy of Sciences on September 6, 1751*

[ *Слово о пользе химии* ]

Speaking about the well-being of human life, one cannot find anything more perfect than when someone brings benefit through pleasant and blameless labors. Nothing on earth can be given to mortals that is higher and nobler than the exercise in which beauty and importance, alleviating the sense of burdensome labor, encourage with some sweetness, delighting the innocent heart and, multiplying the pleasures of others, arouse perfect joy in their gratitude. Such pleasant, blameless, and useful exercise where is it more likely to be found than in learning? In it, the beauty of various things is revealed, and the wonderful diversity of actions and properties, arranged and disposed by divine art and order, is displayed. Enriching oneself does not harm anyone, as it acquires an inexhaustible and universally shared treasure for itself. Those who exert their efforts in it benefit not only themselves but the whole society, and sometimes serve the entire human race. All this, so justly and so much enhancing the happiness of our lives with the ingenuity and efforts of diligent people, clearly demonstrates the condition of European inhabitants, compared with those wandering in the American steppes. Imagine the difference between them in your thoughts. Imagine that one person can only name a few essential things in life that are always present before them, while another not only describes everything that the earth, air, and water produce, not only explains through language all that art has produced over many centuries, but also vividly and clearly portrays concepts that are not subject to our senses. One cannot even count beyond his fingers, while the other not only comprehends magnitude without weight, comprehends quantity without measure, shows from afar the distances of inaccessible things on earth, but also determines the vast distances of celestial bodies, their extensive immensity, their swift motion, and their constantly changing positions at every moment. One cannot show the years of his life or the brief span of his children's lives, while the other not only arranges the various and almost countless adventures of past times, experienced in nature and in societies, by years and months, but also accurately foretells many future events. One, thinking that beyond the forest where he was born, the sky and the earth have merged, regards a fearsome beast or a large tree as the deity of his small world, while the other, imagining the vast expanse, the intricate structure, and the beauty of all creation, reverently regards the infinite wisdom and power of the Creator with a certain sacred awe and reverential love. Imagine a person barely covering his nakedness with leaves or the raw hide of an animal, compared to the one clothed in gold-embroidered garments and adorned with the sparkle of precious stones. Imagine one lifting a stone or a fallen tree from the ground for his defense, armed with bright and sharp weapons and machines imitating thunder and lightning, while another, with the aid of powerful and intricate machines, sets a thin piece of wood against a cumbersome stone, using much effort to move it, to hasten long-lasting tasks, and to accurately measure and divide quantity, weight, and time. Mentally behold one floating across a small river on a tied reed and another sailing through the vast ocean on a great ship fortified with reliable weapons, propelled against its own course by the force of the wind, and instead of a guide, possessing a stone to guide it through the waters. Is it not clear that one is almost elevated above the fate of mortals, while the other barely differs from dumb animals? One enjoys the pleasant glow of clear understanding, while the other, in the dark night of ignorance, can barely see his own existence? Only learning brings great benefit, only illuminates the human mind with bright rays, only the enjoyment of its beauty is pleasant! I would like to introduce you to the magnificent temple of this human prosperity, I would like to show you in it, in detail, through keen insight and tireless reasoning, the brilliant adornments invented by wise and industrious men, I would like to astonish you with their various excellences, to delight you with their abundance, and to attract you to them with invaluable benefit. But to carry out such an undertaking requires more of my understanding, more of my eloquence, more time than is allowed for the fulfillment of this intention. Therefore, I ask you to follow my thoughts into just one inner chamber of this great edifice, in which I will endeavor to briefly show you some of the rich treasures of nature and to announce the use and benefit of those changes and phenomena which chemistry produces in them. If my words are insufficient in demonstrating and explaining them, reward them with the keenness of your own intellect. The knowledge acquired by learning is divided into sciences and crafts. Sciences provide clear understanding of things and reveal the hidden actions and properties of causes; crafts employ them to increase human benefit. Sciences satisfy innate curiosity, while crafts delight in gaining profit. Sciences show the way to crafts; crafts accelerate the origin of sciences. Both serve the common good. How great and how necessary the application of chemistry is in both of these is clearly demonstrated by the study of nature and many useful crafts in human life.

---

[1] Translated to English from the Russian version of Ref. [1] by V.Shiltsev; footnotes and commentary by V.Shiltsev

      When considering natural objects, we find them to possess properties of two kinds. Some we understand clearly and in detail, while others, though clear in our minds, we cannot depict in detail. The first kind includes size, appearance, movement, and position of the whole object, while the second includes color, taste, smell, medicinal powers, and others. The former can be precisely measured through geometry and determined through mechanics; however, the latter cannot be simply employed with such precision because the former are in visible and tangible bodies, while the latter have their basis in the subtlest and most remote particles from our senses. But for an accurate and detailed understanding of any object, one must know the parts that constitute it. For how can we reason about the human body without knowing the composition of bones and structures for its reinforcement, the union, the position of muscles for movement, the spread of nerves for sensation, the arrangement of internals for the preparation of nutritive juices, the extension of veins for blood circulation, and other organs of this marvelous structure? Similarly, detailed knowledge of the second kind of qualities mentioned above is impossible without

investigating the smallest and indivisible particles from which they originate and whose knowledge is necessary for natural investigators, just as these particles are necessary for the composition of bodies. And although in the present age, invented microscopes have greatly enhanced the power of our vision, allowing us to clearly recognize many parts even in the tiniest dust particle, these useful instruments serve only for the examination of organic parts, which are very fine and invisible to the naked eye, constituting the solid parts of animals and growing things, while they cannot present to the sight the particles from which mixed substances originate. For example, through chemistry, it is known that cinnabar contains mercury and vinegar contains white earth, yet neither mercury in cinnabar nor white earth in vinegar can be seen through the best microscopes; they always appear the same. Therefore, knowledge of these matters must be attained only through chemistry. Here I see, you may say, that chemistry shows only the materials from which mixed bodies are composed, but not each of their particles individually. To this I reply that truly up to this time the keen eye of investigators could not penetrate deep into the interior of bodies. But if ever this mystery is revealed, then truly chemistry will be the first leader, the first to unveil the veil of the innermost sanctuary of nature. Mathematicians infer unknown quantities from some known ones. For this purpose, they combine known quantities with unknown ones, subtract, multiply, divide, equalize, transform, transpose, vary, and finally find the sought-after. Following this example, when contemplating the countless and diverse changes that chemistry presents through the mixing and separation of different materials, reason must attain through the most minute inspection, measurement, movement, and position of the original particles constituting mixed bodies. When a loving groom, troubled by love, wishes to know the true inclination of his bride towards him, then, conversing with her, he observes changes in her complexion, the turn of her eyes, and the order of her speech, watches her behavior, courtesy, and cheerfulness, inquires of her attendants who serve her in excitement, attire, outings, and domestic exercises, and thus assures himself accurately of her true state of heart. Similarly, the careful lover of beautiful nature, desiring to experience only the deeply hidden state of the original particles constituting bodies, must scrutinize all their properties and changes, especially those shown by his closest assistant and companion who has access to the innermost chambers of chemistry, and when she combines the separated and dispersed particles from solutions into solid parts and displays various shapes in them, one should inquire of cautious and astute geometry when solid bodies are changed into liquids, liquids into solids, and different kinds of matter are separated and combined, seek advice from precise and intricate mechanics, and when through the merging of liquid substances various colors are produced, one should investigate through penetrating optics. Thus, when chemistry dissects the hidden treasures of its wealthy mistress, the curious and vigilant steward of nature will begin to measure them through geometry, weigh them through mechanics, and scrutinize them through optics. It is highly probable that he will then attain the desired secrets. Here, you may wonder why, up to this time, researchers of natural phenomena have not accomplished so much in this regard? To this, I reply that this requires a chemist of great skill and a mathematician of profound knowledge in one person. A chemist is needed who not only understands this science from reading books alone but who has diligently practiced it with his own skill, and not one who, though making a great number of experiments, hastens solely towards the fulfillment of his desires and, motivated by the desire for great and quickly acquired wealth, rushes only to the realization of his wishes, and, following his dreams, despises the occurrences and changes that happen in his labors, which serve to explain the natural mysteries. Not one is required who is only skilled in difficult calculations, but one who, accustomed to inventing and proving things with mathematical rigor, can derive the hidden truth of nature through precise and unwavering order. Just as useless are the eyes of one who wishes to see the interior of a thing but lacks hands to open it. Equally futile are the hands of one who lacks eyes to observe the open things. Chemistry may rightfully be called the hands, and mathematics the physical eyes. But just as both require assistance in the exploration of the internal properties of physical bodies, so, on the contrary, human minds often diverge into different paths. The chemist, seeing various and often unexpected phenomena and products in every experiment and being attracted to them for the speedy attainment of profit, scoffs at mathematics, which, though it trains in some futile reflections on points and lines, is nevertheless necessary. The mathematician, on the other hand, confident in his clear positions and able to deduce unknown quantities and properties through irrefutable and continuous consequences, despises chemistry, which, burdened only by one practice and misled among many disorderly experiments, is accustomed to pure paper and bright geometric instruments and abhors chemical smoke and ashes. And thus, these two, united for the common good, often give birth to sons of diverse minds. This is the reason why a perfect knowledge of chemistry combined with a deep understanding of mathematics has not yet been achieved. And although in this century some have shown notable success in both sciences, yet this enterprise is considered beyond their capabilities, and for this reason, they do not wish to exert themselves in the investigation of the aforementioned particles with firm intention and steadfast determination. Especially when they have noticed that some, with much labor and time spent, have obscured the natural science with empty thoughts and hallucinations born of mere whims more than they have enlightened it.

   The investigation of the primфкн particles, the constituent bodies, entails the search for the causes of their mutual union, by which they are joined in the composition of bodies and from which arises all the variety of hardness and fluidity, cruelty and softness, flexibility and brittleness. How can all this be better experienced than through chemistry? It alone softens them in fire and again consolidates them, then, dividing them into air, raises them and collects them back, then dilutes them with water and, thickening them in it, firmly connects them, then, dissolving in acidic waters, turns solid matter into liquid, liquid into dust, and dust into stone hardness. And thus, by multiplying and diminishing them in countless ways, the force of mutual cohesion among the parts opens up to the curious physicist a great multitude of different paths, along which he may reach the great art of this cunning nature. But my brief word cannot embrace the mixture and separation of other minerals, as well as growing and animal materials in the changes of the properties of this pleasant body presented to sight. For all these, like some pantomimes or silent interpreters of thoughts on the vast theater of nature, strive to declare their hidden causes to the discerning observer through various changes and, as it were, to interpret them with some speechless conversation.

Animals and growing bodies consist of organic and mixed parts. The mixed ones are either solid or liquid. The liquids contain solids within them, while the solids are nourished by liquids, grow, bloom, and bear fruit. In these processes, nature changes the properties of juices in variously arranged vessels, especially their taste and aroma. It separates sweet milk from bitter bile from the same food and, on the same land, produces sour and pungent fruits and herbs together with fragrant ones. To those who understand the structure of animate bodies and the multitude of earthly phenomena, it is well known that many modifications occur in all these processes. Chemistry endeavors to imitate nature precisely in all these aspects: it often softens strong tastes and refines weak ones; it produces sweetness surpassing that of lead from acetic acid and from sharpness it creates the honey-like sweetness, and through the blending of minerals, it emits the delicate fragrance of pleasant roses. Conversely, from saltpeter, which has no smell or strong taste, it produces penetrating acidity that corrodes solid metals, and a stench that steals one's breath. Is it not clear from this that the search for the cause of different tastes and smells can be undertaken with desired success by following the indications of preceding chemistry and applying its art to discern the hidden changes that are sensitive only to taste and smell in the subtle vessels of organic bodies?

A significant part of physics and the most beneficial science for humanity is medicine, which, through understanding the properties of the human body, identifies the causes of impaired health and, using appropriate remedies for correction, often restores those afflicted by illness from the brink of death. Diseases mostly arise from the disturbance of liquid substances necessary for human life, which, when turned in our bodies, undergo qualities, changes in composition, and beneficial or harmful alterations, methods of which can only be known through chemistry. It is through chemistry that the natural mixture of blood and nutritive juices is understood, the composition of healthy and harmful foods is revealed, and useful medicines are prepared not only from various herbs but also from minerals taken from the depths of the earth. In short, a physician cannot be perfect without sufficient knowledge of chemistry, and all deficiencies, excesses, and corrections arising from them in the medical science should rely on chemistry alone.

It would take long to enumerate and explain in detail what has been discovered through chemistry in nature and what should be discovered in the future. Therefore, I will present to you now only the most important of its actions. Fire, which in its moderate heat is called warmth, spreads widely throughout the world with its presence and action, for there is no place where it is not. Even in the coldest, northern regions near the poles, it always reveals itself in a gentle manner amidst winter. There is no action in nature for which it should not be ascribed as the basis, for all internal movements of bodies, and consequently external ones, occur from it. All animals begin, grow, and move by it. Blood is circulated and our health and life are preserved by its force. Mountains produce various kinds of minerals in their depths and pour out waters to heal the weaknesses of our bodies by its power. And you, pleasant fields and forests, only cover yourselves with beautiful attire, invigorate the members, and delight our senses when the lovely warmth, by its gentle advent, disperses frost and snow, nourishes you with abundant moisture, adorns you with sparkling and fragrant flowers, and enriches you with sweet fruits. Moreover, without fire, nourishing dew and dissolved rain cannot descend upon the fields; without it, springs will be sealed, the flow of rivers will cease, the stagnant air will lose its movement, and the great ocean will freeze into eternal ice; without it, the sun will extinguish, the moon will darken, the stars will disappear, and nature itself must die. Therefore, not only many researchers of internal body mixtures desired to be called by the venerable name of philosophers, not only pagan peoples, among whom sciences were highly esteemed, gave divine honor to fire, but also the sacred scripture repeatedly tells of the divine manifestation in the form of fire. Thus, what is more worthy of our investigation from natural things, such as the soul of all created things, the cause of all wondrous changes, born in the depths of bodies, the subtle and powerful instrument? But this investigation cannot be undertaken without chemistry; for who can know better the properties of fire, measure its power, and open the way to its hidden actions than chemistry, which produces all its effects through fire, using not ordinary means? It suddenly produces fire in cold bodies and great cold in warm ones. It is known to chemists that strong spirits, when dissolved in metals, heat up without the touch of external fire, boil, and emit scorching vapor, that through the fusion of strong nitric acid with certain fatty substances, not only dreadful boiling, smoke, and noise occur, but also a fierce flame arises in an instant. Conversely, warm nitrate, when diluted in warm water, produces such a strong cold that it freezes in a suitable vessel in the middle of summer. I will not mention here the various phosphorus substances invented by chemical art, which ignite in free air and, together with the aforementioned phenomena, clearly demonstrate that the properties of fire can be explored only through chemistry. No one can come closer to this great altar, lit by the supreme from the beginning of the world, than this nearest priestess.

This is the benefit that physics draws from chemistry. This is the method that, through clear knowledge of things, sheds light and shows the direct path to the crafts, in which this science is so indispensable and powerful. I will now try to briefly show this.

Among the crafts, metallurgy holds the first place in my opinion, as it teaches to find and refine metals and other minerals. This advantage gives it not only great antiquity, which isш ndisputable according to the testimony of sacred scripture (* *Book of Genesis*, Chapter 4) and the deeds of humanity, but also an unspeakable and widely spreading benefit. For metals provide strength and beauty to the most important things necessary in society. They adorn the temples of God and shine on royal thrones, they protect us from enemy attacks, they reinforce ships, and bound by their power, they safely navigate between turbulent whirlpools in the sea. Metals open up the earth's depths to fertility; metals serve us in catching terrestrial and marine animals for our sustenance; metals facilitate commerce with convenient coinage, instead of cumbersome and burdensome bartering. In short, no art or craft can avoid the use of metals. However, these necessary materials, especially those of greater value and worth, besides being deeply buried in the earth for our toil, often conceal themselves by their external appearance. Precious metals, when mixed with common earth or combined with lowly stone, elude our sight; on the contrary, simple and even small quantities often sparkle like gold and, with the variety of pleasant colors, entice inexperienced minds to acquire great wealth. And although sometimes an unknowledgeable person accidentally discovers a precious metal in a mountain, yet there is little benefit to them when they cannot separate the valuable from the worthless mixed with it, or when, in separating the majority, they waste it due to their lack of skill. In such cases, how penetrating and powerful is the action of chemistry! Nature in vain hides its treasures from it behind a contemptible veil and locks them in simple chests, for the acumen of chemical fingers can discern the useful from the worthless and the precious from the base, and through the apparent surface, it sees the inner worth. In vain does wealth enclose itself with the unyielding hardness of heavy stones and surround itself with materials harmful to our lives, for armed with water and flame, chemistry breaks the strong rivets and drives away everything contrary to health. In vain does this golden fleece encircle itself with only the trunk of the fierce and fearsome dragon, for the seeker thereof, instructed by our benign Medea, will knock out its poisonous teeth and shield himself from the deadly vapors with the remedies given by her. This benefit from chemistry begins both in our homeland and is fulfilled similarly to what happened in Germany, which the ancient Roman historian Cornelius Tacitus once discussed (* *On Germany*, Chapter 5)[2]. "I cannot say", he wrote, "that silver and gold are not found in Germany, for who has sought them?" And just as great wealth was acquired there in the following centuries, as evidenced by the glorious Meissen and Harzen plants[3], so too can it be expected in Russia, especially since we have not only sufficient experiments but also evident profit in this regard. It is futile to argue that in warm regions more precious metals are born from the action of the sun than in cold ones, for according to reliable physical research, solar heat does not penetrate to such depths in the earth where metals are found. And scorching Libya, devoid of metals, and icy Norway, containing pure silver in its rocks, contradict that opinion. The only difference lies in the fact that there, metals lie closer to the earth's surface, the reasons for which are clearly visible. Firstly, heavy rains often pour there, and in some places, they continue incessantly for half a year, softening and eroding the soil and carrying away light mud, leaving behind heavy minerals; therefore, the inhabitants of those places always search for gold and precious stones in suitable locations after the rainy part of the year. Secondly, frequent earthquakes crush and overturn mountains, and what nature has produced in their depths is thrown to the surface. Thus, it follows that hot regions deprive our northerners of the advantage of acquiring metals with a smaller quantity but with greater ease. But this diligence of northern inhabitants, surpassing those living under the hot belt, should be rewarded. Russia, vast and abundant, requires calculations and efforts for the discovery of metals. It seems to me that I hear her speaking to her sons: "Spread out your hopes and your hands into my bosom, and do not think that your efforts will be in vain. My fields repay the toils of the farmers many times over, and my fertile lands multiply your herds, and my forests and waters are filled with animals for your sustenance; all this not only satisfies my boundaries but also overflows into foreign lands. Therefore, can you imagine that my mountains will not reward the faces of your children with precious treasures? You have in my regions, lying towards warm India and the Frozen Sea, sufficient signs of my underground wealth. To facilitate the conveyance of necessary things for this purpose, I open rivers flowing far in the summer and lay smooth snow in the winter. From your labors, I expect an increase in trade and crafts, I expect the adornment, strengthening, and multiplication of cities, I expect and desire to see my vast seas covered with numerous and formidable fleets, and the glory and power of my state spread across the great ocean to unknown peoples". Be at peace about this, blessed country, be at peace, my dearest homeland, when within you reigns a generous patroness of learning. The great enlightener[4] has sought and multiplied solid metals for your defense; his august daughter seeks and multiplies the precious ones for your adornment and enrichment, spreading along with other sciences the art of chemistry, which, nurtured by the care of this great monarch, fortified and encouraged by her generosity, will penetrate into the heart of the mountains and, discovering what lies within them unused, will purify it for the multiplication of our bliss, and beyond this, with its strong action in metallurgy, it will strive to bring other useful fruits to you.

---

[2] Publius Cornelius Tacitus, "De Origine, situ, moribus ac populis Germanorum", chapter 5; it's a historical, geographic and ethnographic work on the Germanic tribes outside the Roman Empire.
[3] Reference to metallurgic plants in Meissen (Saxony) and in Harz mountains in Germany.

So widely extends Chemistry its hands into human affairs, listeners. Wherever we look, wherever we turn, we see the successes of its diligence before our eyes. From the earliest times of the world's creation, the heat and cold compelled man to cover his body; then, from the first use of leaves and animal skins, he conceived of making clothes from wool and other soft materials, which, although they served well enough to protect his body, yet the human heart, weary of one appearance, and its fickle desire for change, demanded variety, envied the variegated fields, and sought similar splendor even in clothing. Then chemistry, extracting juices from herbs and flowers, boiling roots, dissolving minerals, and combining them in various ways, sought to fulfill human desires, and how much it has adorned us, you do not need my words to prove, but with your own eyes you always clearly see it.

These chemical inventions not only delight our eyes with changes in clothing but also satisfy our other inclinations. What better arouses greater diligence in ourselves and respect than our parents? What is dearer to a person in life than his own children? What is more pleasant than sincere friends? But their absence in distant places or their departure from the light takes them out of our sight. In such a state, what can comfort us more and soften our heartfelt sorrow than the likeness of their faces depicted by the art of painting? It presents the absent as present and represents the departed as living. Everything that has been removed from our sight by the length of time or the distance of place, painting brings near and exposes to it. With it, we see the great rulers and brave heroes who preceded us, and other great people deserving of glory among descendants. We see distant cities in far-off lands and magnificent and huge buildings. When wandering in vast fields or among high mountains, we gaze at the tossing sea, at the ships being crushed or running to shelter with capable breezes. In the midst of winter, we delight in the sight of green forests, flowing springs, grazing herds, and toiling farmers. All this we owe to painting.

---

[4] Peter the Great

But its perfection depends on chemistry. Take away the colors invented by her artisanship, and we lose the pleasantness of the painted images, the similarity with things, and even their liveliness, which we derive from them. It is true that colors do not retain their clarity and goodness for as long as we wish, but change in a short time, darken, and eventually lose a great part of their beauty. To prevent this disadvantage, where could one resort? Who could invent a means for the long-term and inevitable permanence of paintings? The same chemistry, which, seeing that her delicate compositions wither and deteriorate from the strict changes of the air and the rays of the sun, employed the strongest instrument of her artisanship, fire, and, combining solid minerals with glass in great heat, produced materials that surpass the brightness and purity of previous ones in workmanship, and oppose themselves to the humidity of the air and the heat of the sun so strongly that through many centuries they have lost none of their beauty, as evidenced by the temples erected thousands of years ago in Greece and Italy and beautified by mosaics. And though in ancient times natural stones of various colors were sometimes used for this purpose, because then, and even in ordinary painting, natural earths served as compositions for lack of colors invented by art, yet the great advantages which glass compositions have over stones have attracted skilled Roman artists to their use in the present time. Firstly, it is rare and very difficult to obtain shadows of many colors from natural stones, which come out in compositions at the discretion of the artist. Secondly, although sometimes with great difficulty they are obtained, yet they must spoil not a few expensive stones suitable for other purposes. Thirdly, from compositions, for their greater softness, parts of the desired size and shape can be separated and melted, which natural stones require much effort and patience. Finally, through the craft of chemistry, glass painted with the richness of colors akin to natural stones has been crafted, and henceforth, through the diligence of chemists, even greater perfection can be achieved.It is true that stones surpass glass in hardness, but in this matter, it is useless, where only the constancy of colors in sunlight and air is required. So, not in vain do modern masters in this art prefer the art to nature, which produces better effects with less effort and expense. Proposing this sole use of glass in the art of painting, I can hardly refrain from briefly showing many other benefits derived from this great chemical invention. However, the presentation of it requires a separate discussion, which is not appropriate for my current endeavor. Therefore, I hasten to other activities of our science, which demonstrate its power in the arts. And what a vast expanse I see before me! There are still various matters that attract my attention, each one more than the last. And when I wish to present to you the many ways chemistry assists us in preparing pleasant foods and beverages, I must first discuss the vessels from which we enjoy them. Imagine their cleanliness, transparency, brilliance, and various decorations, which enhance the pleasure of tasting with their artistry, delighting both the tongue and the eye. Therefore, I do not wish to exhaust your patience with a detailed enumeration of everything, but I will conclude with a single, salvific boon bestowed upon humanity by chemistry.

    If we contemplate the lamentable adventures and changes in ancient times in various countries, how often do we read with pity in histories of the sudden invasion of distant and unknown peoples, the transformation of great and glorious cities into smoke and ashes, the

devastation of villages and entire nations who could not resist the swift enemy, resulting in ultimate ruin and dispersal, so that only the name remains of their former might and glory. They tell of fields filled with the bodies of many thousands and wide rivers flowing with blood and strewn with corpses, surpassing the likelihood of our times, in which we have only a few terrible examples. However, the gravity of such renowned writers and the very ruins of ancient cities remove all doubt about the justice of these tearful spectacles. Whence, then, has come the moderation that has settled among mortals? Was it not some Orpheus who softened human passions with sweet song? Yet even in our times, hearts torn by envious rivalry seek to seize the possessions of others. Did not Lycurgus or Solon bind passions with strict laws? Yet today, the power of arms is often valued more than the rights of the people. Did not the great Cyrus, with his name alone, satisfy greedy avarice many times over? Yet this is like a flame that, the more fuel it receives, the more fiercely it burns. Who then bestowed such a great boon upon us? Who mitigated such fierce bloodshed? A simple and poor man, who, fleeing his poverty, followed the path of chemistry to obtain wealth by unknown means, intending to gain access to precious metals, mixed sulfur and saltpeter with coal and placed them in a vessel over fire. Suddenly, a terrible sound and a powerful blow followed! And though he himself was not without injury, he was more than compensated by the hope of obtaining a strong and indestructible metal-breaking substance. He sealed and riveted his composition in solid iron vessels, but without success. Hence arose firearms; battlements and city walls resounded, and from human hands a deadly lightning bolt flashed! But you may ask, does this not kill instead of giving life, and does it not achieve more harm than good? I answer: the more it destroys, the more it saves. Consider a battle in which warrior fights against warrior, sword against sword, blow against blow, in close combat; must not in a single instant many thousands fall dead or mortally wounded? Compare this with the battles of today, and you will see that it is easier to raise a hand than to load a gun with powder and shot; it is easier to strike an enemy within reach on clear ground than to aim at a distant target through dense smoke, shaking hands from the glare and the heat of the sun; the heart burns brighter against an adversary whom one can see directly facing oneself, rather than one who is concealed. This is the reason why there are no Hannibals or their like in our times, who, like him, removed golden rings from the fingers of slain Roman nobles by the handful in a single battle. There are no inhuman Batu Khans, who, in a short time, traversed from the Caucasus to the Alps, laying many lands to waste. The sudden enemy dare not disturb the peace of resting nations now, but fears losing not only its spoils but also its life, leaving behind fortified strongholds, equipped with this new invention, not only to plunder but also to protect them. On the contrary, whoever has the strength to destroy such fortifications with the aid of this chemical invention cannot easily reach distant places; an army burdened with heavy projectiles cannot compare with the swift report, which announces impending disaster and gathers peoples for their defense. Thus, chemistry has diminished human calamity with its most powerful weapon and saved many from death by death! Rejoice, uninhabited places, and adorn yourselves, impassable deserts: your prosperity is approaching. Tribes and nations are multiplying visibly, and they are spreading more quickly than before; soon, great cities and abundant villages will adorn you; instead of the roar of wild beasts, your spaces will be filled with the sight of rejoicing humans, and instead of thorns, they will be covered with wheat. But do not forget to thank chemistry, the great participant in your well-being, which desires nothing from you but diligent practice in it, for your greater adornment and enrichment.

When presenting the benefits of chemistry in both sciences and crafts, listeners must be warned not to assume that all human well-being is solely reliant on this discipline, and that I, with some imprudent enthusiasts of my own profession, disdainfully regard other sciences. Each science contributes equally to our happiness, as you heard earlier in my discourse. Humanity must offer profound gratitude to the Almighty for the ability bestowed upon it to acquire such knowledge. Europe, which above all nations enjoys these gifts and thereby distinguishes itself from others, should contribute even more. However, Russia should sacrifice on the altar of progress with fervent zeal. At a time when sciences emerged from the darkness of barbaric ages, Russia, under the courageous hand of the wise hero, Peter the Great, the true father of the fatherland, embraced enlightenment. Despite facing internal and external adversaries, Peter, endowed with divine strength, overcame all obstacles and paved the way for clear understanding. After the arduous military endeavors and ensuring the security of the entire country, his first concern was to establish, consolidate, and promote sciences within it. Blessed are those who beheld this divine man on earth! Blessed are those who shed their sweat and blood with him for him and for the fatherland, and whom he embraced with his lips anointed with his kisses as a reward for faithful service. But for us, who did not have the privilege of witnessing this great ruler in life, we now have strong solace in seeing his worthy daughter and heiress on the throne, our most gracious autocrat. We see in her a pious daughter of a godly father, a courageous daughter of a hero, a discerning daughter of a wise man, a generous protector of their legacy. The sciences see how she regards them with maternal care and with reverent diligence, and they wish that during her blessed life and prosperous reign, not only this assembly but the entire fatherland would be pleased with their learned sons.

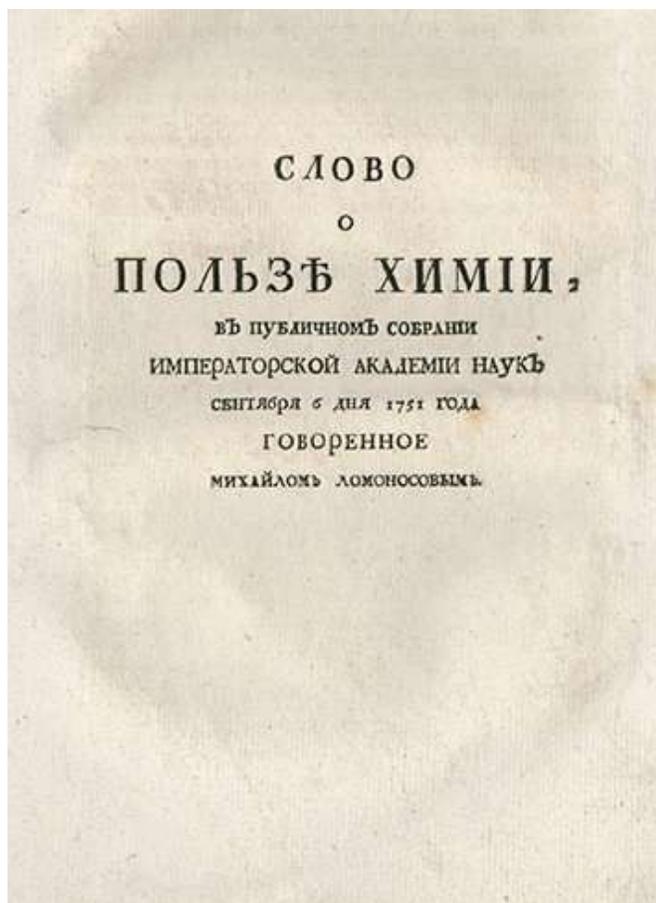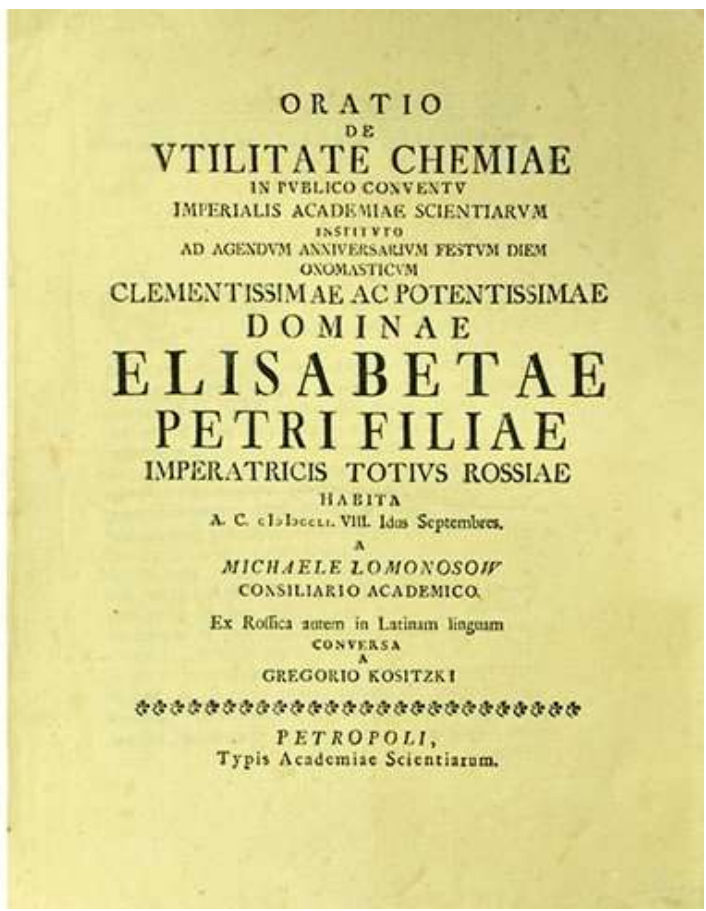

The title pages of "Speech on the Benefits of Chemistry at a Public Meeting of the Imperial Academy of Sciences, September 6, 1751, Spoken by Mikhail Lomonosov", published in Russian (1751) and Latin (1759). The Russian National Library (Saint Petersburg).

*Commentary* [by V.Shiltsev]:

This English translation of Mikhail Lomonosov's seminal work is derived from its Russian [1] (also Ref.[2], v.2, pp.345-369). It continues the series of English translations of Lomonosov's nine most important scientific works, which were included by Lomonosov himself in the convolute *Lomonosow Opera Academica* and sent for distribution among scientific Academies in Europe. Five other translations of these publications can be found in Refs. [3-7]. Henry Leicester's book [8] includes translations of "Oration on the Use of Chemistry" together with "Oration on the Origin of Light" and "Meditations on the Solidity and Fluidity of Bodies". Unfortunately, these translations are found to be not fully satisfactory as on numerous instances the meaning of the Russian original was lost or grossly distorted, mostly due to very difficult 18th Russian of the Lomonosov's original publication, which is not always easily readable and understood even by the Russian speakers of nowadays. All that makes Lomonosov's papers hard to translate to English. This text is free of the deficiencies of the H.Leicester's translation (that was used only for reference) and was initially composed with the aid of Google Translate and ChatGPT programs, which provided a rough draft from both Russian and Latin versions of the Lomonosov's "*Oration…*". However, the machine translations were found to be unsatisfactory as they often failed to capture the nuances and complexities of the original text - mostly due to very difficult old Russian language of the original. Therefore, extensive reworking and improvements by the translator were indispensable to ensure the translation accurately conveys the intended meaning and is free from semantic deficiencies. More on the life and works of the outstanding Russian polymath and one of the giants of the European Enlightenment can be found in books [9, 10] and recent articles [11-18].

The history of this "*Oration…*" is as follows: On February 20, 1749, the academic Chancellery presented to the President of the Academy of Sciences, count K. G. Razumovsky, the program of a public assembly of the Academy, where it was proposed to assign the delivery of a speech on the benefits of chemistry to Lomonosov, "who is capable of writing a dissertation both in Russian and in Latin, and either publicly reading it or speaking it by heart" ([19], p. 122). On September 6, 1751, Lomonosov delivered the "Oration on the Benefits of Chemistry" at the public assembly of the Academy. It was immediately printed in the collection of speeches delivered at this meeting. In 1758, the "Word on the Benefits of Chemistry" was translated into Latin by G. V. Kozitsky and published as a separate edition in the same year (see [1]).

In the "*Oration…*" Lomonosov gave an elaborate exposition of his views on the tasks and significance of chemistry for the development of industry and its path in our country, and dwelled on his idea of the inseparable connection between scientific theory and practical activity, a thought that permeates all his work.Defining chemistry as a science developed in close connection with physics and mathematics, Lomonosov further links the future successes of chemistry with his doctrine of "insensitive" physical particles. In the "Oration on the Benefits of Chemistry," many fundamental principles of his early works - such as, e.g., "Elements of Mathematical Chemistry" ([2], v.1, pp.65-83, and "An Attempt at a Theory of Insensible Particles of Bodies and of the Causes of Particular Qualities in General" ([2], v.1, pp.169-235), and some others – were further developed. At the same time, the "Word on the Benefits of Chemistry" outlines the program of Lomonosov's physico-chemical research (see. e.g., [18]) and presents some results of his experimental work.

The "Oration on the Benefits of Chemistry" was widely known to both Lomonosov's contemporaries and subsequent generations in Russia. It is a splendid example of Lomonosov's oratorical prose, receiving fair and high praise from such an eminent connoisseur and master of the Russian literary language as the poet K. N. Batiushkov (see: *Works of K. N. Batiushkov*, vol. II, St. Petersburg, 1885, pp. 344-346). Lomonosov himself highly valued this speech. Besides including it in the *Opera Academica* convolute, he lists this "*Oration…*" in his 1751 annual report 1751 (PSS [2], v.10, p. 511) and in the "Review of the most important discoveries with which Mikhailo Lomonosov has tried to enrich the natural sciences" which he compiled in 1764 (PSS [2], vol. 10, p. 404-411).

I would like to thank Prof. Robert Crease of SUNY, my long-term collaborator and co-author of several scholar papers on Mikhail Lomonosov, for the encouragement to translate Lomonosov's major works to English.